\documentclass{MITcsail}

\definecolor{darkred}{RGB}{140, 21, 21}
\definecolor{citegray}{gray}{0.7}
\definecolor{orange}{HTML}{F58025}

\definecolor{deepred}{rgb}{0.631,0.102,0.102}
\definecolor{amethyst}{rgb}{0.6, 0.4, 0.8}
\definecolor{darkgreen}{rgb}{0.3,0.7,0.3}
\definecolor{salmon}{RGB}{241, 150, 141}
\definecolor{mildyellow}{HTML}{FFF2CC}

\usepackage{xspace}
\usepackage{xcolor}
\usepackage{setspace}

\definecolor{aiblue}{RGB}{66, 133, 244}
\definecolor{humangreen}{RGB}{15, 157, 88}
\definecolor{lightgray}{RGB}{245, 245, 245}
\definecolor{codebg}{RGB}{240, 240, 240}

\newcommand{\airesponsetitle}{}

\newcounter{aimessage}

\newcommand{\humanprompttitle}{}

\newcounter{humanmessage}

\usepackage{cleveref}
\usepackage{float}
\usepackage{array}
\usepackage{makecell}  
\usepackage{booktabs}  
\newcolumntype{P}[1]{>{\raggedright\arraybackslash}p{#1}}   
\usepackage[T1]{fontenc} 
\usepackage{hyperref}
\usepackage{tikz, xurl}
\usepackage{xcolor}
\usepackage[most]{tcolorbox}
\usepackage{arydshln}
\usepackage{enumitem}

\hypersetup{
    colorlinks=true,
    linkcolor=orange,
    citecolor=citegray,
    filecolor=darkred,
    urlcolor=orange,
}
\usepackage[numbers, sort&compress]{natbib}
\usepackage{supertabular}

\title{How Can AI Augment Access to Justice? Public Defenders' Perspectives on AI Adoption}

\author{
Inyoung Cheong$^{*,\dagger}$,
Patty Liu$^{*}$,
Dominik Stammbach$^{*}$,
Peter Henderson \\
\normalfont{\small Princeton University} \\
\vspace{1em}
}

\vspace{1em}

\begin{document}

\maketitle
\thispagestyle{firstpagestyle} 

\renewcommand\thefootnote{}\footnote{
\noindent
$^{*}$ Equal contributions, authors listed in alphabetical order.\\
\vspace{0.5em}
\hspace*{1.5em}$^{\dagger}$ Correspondance to: iycheong@princeton.edu
}

\begin{abstract}

Public defenders are asked to do more with less: representing clients deserving of adequate counsel while facing overwhelming caseloads and scarce resources. Although artificial intelligence (AI) is often promoted as a means of relieving administrative and cognitive burdens, legal AI research rarely engages with the everyday realities of public defense work. Drawing on in-depth, semi-structured interviews with 17 public defense professionals across the United States, we identify work-intensive tasks most amenable to AI assistance and the ethical constraints involved in legal representation. We develop a comprehensive task-level map of public defense work, dividing it into five pillars to clarify where AI can and cannot contribute: evidence investigation, legal research \& writing, client communication \& support, courtroom representation, and defense strategies. Interviewees consistently identified evidence investigation, such as reviewing large volumes of digital records, as the area with the greatest potential for AI support. AI was viewed as having more limited roles in legal research and client communication, and as least compatible with courtroom representation and defense strategy. We find that AI adoption is constrained by costs, restrictive office norms, confidentiality risks, and unsatisfactory tool quality. Our interviewees emphasize safeguards for responsible use, including mandatory human verification, limits on over-reliance, and the preservation of relational aspects of lawyering. Building on these findings, we outline a research agenda that promotes equitable access to justice by prioritizing open science, building domain-specific datasets and evaluation, and incorporating frontline practitioners' perspectives into system development.

\end{abstract}

\section{Introduction}\label{sec:01-intro}

In the United States, public defenders provide representation to defendants who cannot afford private counsel. 
However, public defenders are systematically overworked and under-resourced \citep{Pace2023}. This exacerbates disparities in the justice system: underprivileged groups are disproportionately harmed when they receive inadequate representation \citep{nij2023_gideon60, donohue2024_crisis_public_defenders, LegalServicesCorporation2022}. Meanwhile, Artificial Intelligence (AI) and Large Language Models (LLMs) are reshaping work across industries~\cite{mosch2022medical, ma2024integrating, perron2025moving}. General-purpose tools, alongside specialized applications, promise to automate or accelerate legal tasks. Scholars have suggested that AI could help close the justice gap, by developing specialized solutions that assist public defenders \cite{mahari-etal-2023-law}. However, apart from anecdotal evidence about commercial vendors and some public defender offices experimenting with AI tools, little is known publicly about how public defenders themselves view the use of AI in their work. 

\begin{figure*}[t!]
    \centering
    \includegraphics[width=0.96\linewidth, trim=0in 2in 0in 2in, clip]{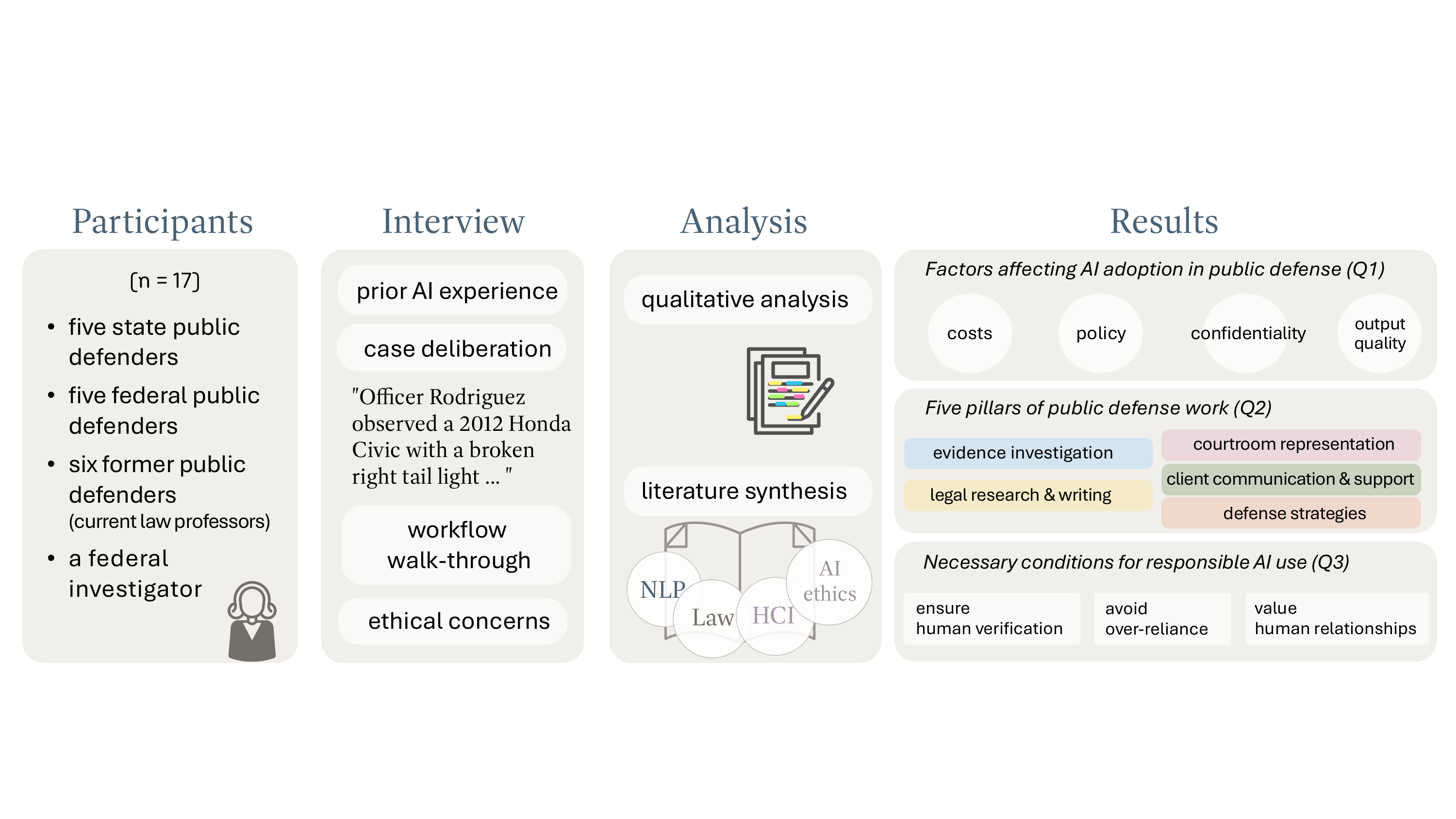}
    \caption{Overview of our research process: we interviewed 17 legal practitioners in public defense on their AI experience, workflow integration, and ethical concerns, using realistic cases as prompts. We applied qualitative analysis and literature synthesis, categorized public defense work into five pillars, examined normative and practical constraints, and extracted principles for responsible AI adoption.}
    \label{fig:overview_methods}
\end{figure*}

To address this gap, we design a qualitative study to elicit the technology demands and hesitations of public defenders. We summarize our methodology in Figure \ref{fig:overview_methods}. Our study addresses three research questions:

\begin{itemize}[nosep]
\item RQ1: What technical, organizational, and cultural factors influence the adoption of AI in public defense?
\item RQ2: Which areas of public defense work present opportunities or limitations for AI adoption?
\item RQ3: How can we ensure responsible AI use in public defense to uphold professional ethics and quality of work? 
\end{itemize}

To answer these questions, we conduct a qualitative analysis of semi-structured interviews with 17 participants who have substantial experience in public defense. Across the board, AI adoption is constrained by four barriers: costs, restrictive office norms, confidentiality risks, and unsatisfactory tool quality. We find that public defenders see the greatest promise in \textit{evidence investigation}, where AI could transcribe, summarize, and synthesize large volumes of digital records. They assign narrower roles to \textit{legal research and writing}, which requires strict manual verification, and to \textit{client communication and support}, where AI may assist with clarity. \textit{Courtroom representation} and \textit{defense strategies} are viewed as least compatible with AI assistance, as they rest on contextual judgment and human relationships. Public defenders also stress safeguards for responsible use, including mandatory human verification, explicit limits on over-reliance, and preserving the relational dimensions of lawyering. 

This study advances the literature in several ways. It centers the voices of public defenders and identifies the tasks they most want supported by AI, aligning with broader calls to ground AI development in the actual needs of frontline workers~\cite{shao2025future}. It introduces a task-level map of public defense through five pillars of practice, challenging simplistic claims that the use of generative AI signals attorney incompetence or that every output must be manually verified. Finally, it outlines a forward-looking research agenda for access to justice, highlighting the potential of open-source models to reduce vendor dependence while ensuring confidentiality and cost-efficiency. Building such systems requires sustained effort from research communities, including the development of domain-specific datasets, evaluation benchmarks rooted in real public defense practice, and participatory design with public defenders and other stakeholders.

\section{Background and Related Work}\label{sec:02-background}

This section situates our study within the broader legal and technical literature. We outline the structure and challenges of U.S. public defender systems and review the emergence of AI in criminal defense, highlighting both its applications and the risks. Finally, we draw on prior scholarship examining how professionals adopt AI in high-stakes domains to illuminate the need for empirical research on public defenders.

\vspace{0.5em}
\noindent\textbf{Public Defender Systems.} 
The U.S. established public defender systems to ensure legal representation for indigent criminal defendants~\cite{wice2005public}. After \textit{Powell v. Alabama} sparked the nationwide concern that defendants facing the death penalty received almost no legal representation in 1932~\cite{powell1932, wice2005public}, eventually in 1962, the U.S. Supreme Court declared the constitutional right to counsel in \textit{Gideon v. Wainwright}~\cite{gideon1963}. In response to Gideon's petition, the Court declared the state's obligation to provide indigent (poor) defendants an attorney. Justice Hugo Black argued: ``[I]n our adversary system of criminal justice, any person hauled into court, who is too poor to hire a lawyer, cannot be assured a fair trial unless counsel is provided for him. \ldots [L]awyers in criminal courts are necessities, not luxuries.''

The constitutional guarantee is not a guarantee of resources, and public defenders in many states struggle: time scarcity, resource limitations, and overwhelming caseloads limit their ability to provide adequate legal representation~\cite{stuntz1997uneasy, ogletree1992beyond, wilkerson1972public, Pace2023}. Utah's public defenders were assigned as many as 525 misdemeanor cases per person, and some Louisiana public defenders have caseloads as high as 1,000 clients. This only allows spending a few hours per client~\cite{staffing}. New Mexico Chief Public Defender was held in contempt, after his office failed to appear in five cases due to the caseload strains~\cite{newmexico}. Excessive caseloads undermine the quality of representation for indigent defendants~\cite{hoffman2005empirical, magana2023public, wice2005public}. 

These challenges manifest differently across states: In some, public defender offices employ full-time attorneys and staff dedicated exclusively to representing indigent defendants. In others, assigned or contracted counsel consists of private attorneys who take cases through court appointments, government contracts, or nonprofit-managed programs
~\cite{washingtonpublicdefender, texaspublicdefener}. Our work asks whether AI technologies can help bridge these systemic resource constraints in indigent defense.

\vspace{0.5em}
\noindent\textbf{AI Adoptions and Concerns in Criminal Defense.}
The emergence of LLMs capable of interpreting formal legal language~\cite{translating_legalese} and document summarization~\cite{zhong2022computing, moro2024multi} has significantly broadened AI's accessibility and application in legal practice~\cite{ariai2024natural}. The capacity of LLMs expanded both the depth and breadth of AI tools in criminal defense~\cite{berkeley}. There are two categories of LLM-based tools. \textbf{General-purpose AI tools} (e.g., ChatGPT, Claude, Gemini) support broad tasks like research, brainstorming, general information seeking, drafting, and summarization, while \textbf{specialized AI tools} (e.g., Westlaw AI, Lexis AI, discovery platforms) are designed for specific legal functions such as case analysis, evidence review, and transcription~\cite{integratingai}. These tools promise efficiency but also raise concerns such as hallucinations, bias, and breaches of confidentiality. These concerns are especially acute for public defenders, who serve as the last line of defense for indigent clients and handle their most intimate data, from jail calls to medical records. 

\begin{itemize}
\item\textbf{Hallucinations.}\label{subsec:lit-concerns}

As LLMs are trained to predict the most likely next token given the context, they are good at producing plausible-sounding, but not factually grounded answers~\cite{magesh2025hallucination}. Mitigating hallucinations is one of the most lively research fields~\cite{forbes_hallucinations, o3o4systemcard}, such as Retrieval-Augmented Generation methods~\cite{lewis2020retrieval, rangan2024fine}.  
Researchers find that Lexis AI and Westlaw AI produce incorrect information 17–34\% of the time, compared to GPT-4's 43\%~\cite{magesh2025hallucination}. As of April 2026, one tracker has documented 1345 incidents of fabricated citations identified by courts, involving 509 attorneys, 803 pro se litigants (individuals representing themselves without attorneys), and 16 judges~\cite{damientracker}. Courts have expressed mounting frustration with the flood of fabricated cases. As one judge remarked, fabricated citations ``have required this court to spend excessive time,'' at the expense of taxpayers and litigants ``many of whom wait years for a resolution.''~\cite{california_hallucinations} Judges nationwide have issued more than 100 orders, ranging from bans on AI-generated filings to mandates requiring disclosure of AI use~\cite{tracker}. The American Bar Association (ABA)'s Formal Opinion 512 emphasizes that lawyers must be aware of the risks of generative AI, verify all outputs, and maintain supervisory responsibilities over how such tools are used~\cite{ABAopinion}.

\item\textbf{Bias.} Bias in AI outputs has raised concerns in criminal justice for their use in predictive policing, and criminal risk assessment~\cite{berk2017impact, berk2019machine, tolan2019machine, propublica, ABA_bias}.  
LLMs are well-documented to replicate and reinforce the biases present in their training data~\cite{dai2024bias, hamidieh2024identifying, hong2025common}. For example, vision-language models associate ``white'' with globalist, ``Middle Eastern'' with terrorist, and ``East Asian'' with authoritarian~\cite{hamidieh2024identifying}. Another study found that vision models generate sexualized images 73\% of the time for a ``17-year-old girl'' prompt, but less than 9\% for comparable male prompts~\cite{wolfe2023contrastive}. Language models show ``dialect prejudice,'' such as recommending harsher sentences, including the death penalty, for speakers of African American English~\cite{hoffman2005empirical}. Such biases could distort narratives when LLMs are applied to public defense tasks, such as transcribing jail calls in African American English, automating case intake processes, conducting forensic analysis of devices seized for alleged possession of inappropriate material, or facilitating client communications where dialect, tone, or expression may be unfairly judged.

\item\textbf{Confidentiality.} 
LLMs and LLM-integrated services accessed through APIs present inherent security vulnerabilities~\cite{evertz2024whispers}. The centralized accumulation of vast amounts of data creates opportunities for adversaries to exploit techniques such as model inversion, membership inference, or extraction attacks to obtain sensitive information~\cite{RajivWeginwar2025MitigatingDL, feretzakis2024privacy, wu2024unveiling}. 
Another concern arises from the practice of many LLM providers using user records to retrain or fine-tune their models. For instance, OpenAI discloses that ChatGPT records may be used for further training~\cite{openai_privacy}. Given LLMs' tendency to ``memorize'' data, this raises the risk of inadvertent storage and reproduction of sensitive information from training data~\cite{liu2023mitigating}. Data preserved by the AI or cloud service providers may fall outside attorney–client privilege or attorney work product, exposing them to disclosure upon civil subpoenas, criminal discovery orders, search warrants, or national security letters~\cite{privilege, cheong2024not}. 
\end{itemize}

\noindent\textbf{Professionals' Perspectives on AI Adoption.}
AI adoption in high-stakes domains has been widely examined in light of both its potential and its pitfalls. A growing body of research emphasizes the importance of hearing directly from professionals who experience technological needs in practice and whose work is guided by professional ethics~\cite{fukumura2021worker, wang2025stakeholder}. Qualitative interviews have explored physicians~\cite{mosch2022medical}, home care workers~\cite{solano2025running}, and mental health professionals~\cite{ma2024integrating}. By interviewing legal professionals, studies surface the conditions under which AI should provide legal advice to the public~\cite{cheong2024not} and legal professionals' trust in such systems~\cite{kennedy2025law}. There are qualitative studies particularly on public defenders, but they examine contested computational forensic tools~\cite{beyond_reasonable_doubt} or the impact of body-worn camera footage on adjudication~\cite{robertson2024body}. Little is known about public defenders' broader perceptions of AI technologies, particularly in the wake of LLMs. This study examines how public defenders view AI's potential to support chronically understaffed offices, the work-intensive tasks most amenable to AI assistance, and the ethical considerations involved in legal representation.

\section{Methods}\label{sec:03-methods}

From March 2025 to March 2026, we conducted 14 semi-structured, in-depth interviews with 17 participants—16 current or former public defenders (including six current law professors) and one investigator working in a public defender office (\textit{see} Appendix~\ref{appendix:participants})—to understand their perceptions of AI. This study is approved by the Princeton University Institutional Review Board (IRB\# 17609). We widely advertised the study through institutional social media accounts, newsletters, listservs, Slack channels, and personal networks.\footnote{\url{https://sites.google.com/princeton.edu/ai-for-public-defenders}} We also conducted targeted outreach to over 150 attorneys who self-identified as having public defense experience via personalized emails and LinkedIn messages. Despite these efforts, response rates were low. One contributing factor was the inability to offer compensation to many participants due to institutional restrictions: our university prohibits compensating government employees, and many public defender offices prohibit accepting compensation. Prior research faced similar challenges. A recent FAccT 2025 study secured 14 participants only after outreach to tens of thousands of researchers due to the constraints in compensation~\cite{youcannot}. Another study on public defenders' perceptions of forensic technology interviewed 13 public defenders and four technologists~\cite{jin2024beyond}.

We reached out to leadership at federal and state public defender offices, hoping institutional collaboration could justify participation time. While management frequently expressed support for the research, only one federal public defender office facilitated actual interviews and others did not materialize due to work overload and scheduling constraints. We recruited assigned and contract-based counsel (not employed by public defender offices) and law professors directing criminal defense clinics, to whom we were able to offer \$50 as compensation for one-hour interviews. Law professors proved particularly responsive, understanding the significance of this research topic. Five of six law professors currently supervise students working in public defender offices and maintain close operational relationships with those offices; one recently retired from a state office and works on criminal justice reform. These participants contributed perspectives spanning multiple jurisdictions and organizational structures. 

Although our sample size aligns with qualitative research norms in HCI~\cite{localstandards}, broader participation would offer additional insights, particularly given substantial variation across federal and state systems and local jurisdictions. Nevertheless, we achieved a balanced pool across jurisdictions (federal, state, county) and employment types (office-employed, contract-based), which enabled cross-contextual and analytically transferable insights rather than exhaustive coverage~\cite{kvale1996interviews}. Most interviews were conducted one-on-one, with two sessions having two participants. The interviews had two parts: (1) case-based reasoning exercises and (2) open-ended questions. We presented realistic fact patterns commonly encountered by public defenders (an example appears in Appendix~\ref{app:scenario}) and observed how participants identified legal issues and developed defense strategies. This method is widely used to elicit experience-based knowledge among professionals with clinical experience~\cite{cheong2024not, feng2023case, Nilsson2009ProfessionalAI, Klein1989CriticalDM}. 

Our interview protocol evolved iteratively as early findings influenced subsequent data collection, consistent with the emergent nature of qualitative methods~\cite{adeoye2021research, Schulze2020LeadershipFO, harper2011qualitative, agan2021your}. Informed by current legal AI literature, we initially centered interviews on legal research assistance. However, early interviews revealed that legal research constituted only part of public defense work. This prompted us to adopt a broader view of the work pipeline and develop more detailed questions about workflow, professional identity, and attorney–client relationships. By approximately the 10th participant, we reached ``theoretical saturation,'' at which point additional interviews no longer altered the conceptual structure of the phenomenon~\cite{saunders2018saturation}. The five core domains of public defense work had stabilized; subsequent interviews added nuance rather than fundamentally changing the structure. After incorporating additional inputs from seven more participants, we considered the sample sufficient for the conceptual objectives of this study, while acknowledging that broader geographic and institutional variation remains an important direction for future work. Appendix~\ref{appendix:interview_evolution} documents the evolution of our interview protocol. 

All interviews were conducted via Zoom with participants' consent. We generated automated transcriptions, followed by manual corrections. We analyzed 14 transcripts, with a total word count of 84,908. One author cleaned transcripts to preserve conversational tone while enhancing readability. Three authors conducted abductive coding manually~\cite{tavory2014abductive, cheong2024not}, using an open-source analysis tool Taguette~\cite{rampin2021taguette}. Consistent with grounded theory methods, we used emergent themes from data collection to concurrently refine our interview questions and inform our analytical framework. Our findings enter into dialogue with literature on AI integration in legal professions~\cite{bhattacharya2021incorporating, frankenreiter2022natural, translating_legalese, almeida2024legal, nithya2024ai} and its normative and practical concerns~\cite{legg2019artificial, fagan2020natural, cheong2024not}. Three authors engaged in iterative codebook refinement through multiple team meetings, ensuring that all transcripts were assigned to at least two coders. Through these meetings, multiple coders examined all documents and reached consensus on codes through discussion, rendering inter-rater reliability metrics unnecessary~\cite{mcdonald2019reliability}.

\section{Results}\label{sec:04-results}

Our main results consist of insights on how AI tools can potentially benefit public defenders. All participants have some experience using AI tools, and several attended AI training. However, most participants (59\%) stated they had not used AI for professional purposes. We first discuss barriers to wider adoption, followed by a categorization of public defenders’ tasks summarizing their workflow and identifying opportunities for AI assistance. We conclude with key ethical considerations participants raised around AI usage.

\subsection{Barriers Preventing Adoption (RQ1)}

We identify four primary reasons for limited adoption of AI tools: (1) prohibitive costs especially for self-employed attorneys, (2) office norms and policies, (3) unresolved concerns around confidentiality, and (4) current tools' unsatisfactory results. 

\begin{figure*}[t!]
    \centering
    \includegraphics[width=0.90\linewidth, trim=1.5in 1.5in 1.5in 1.8in]{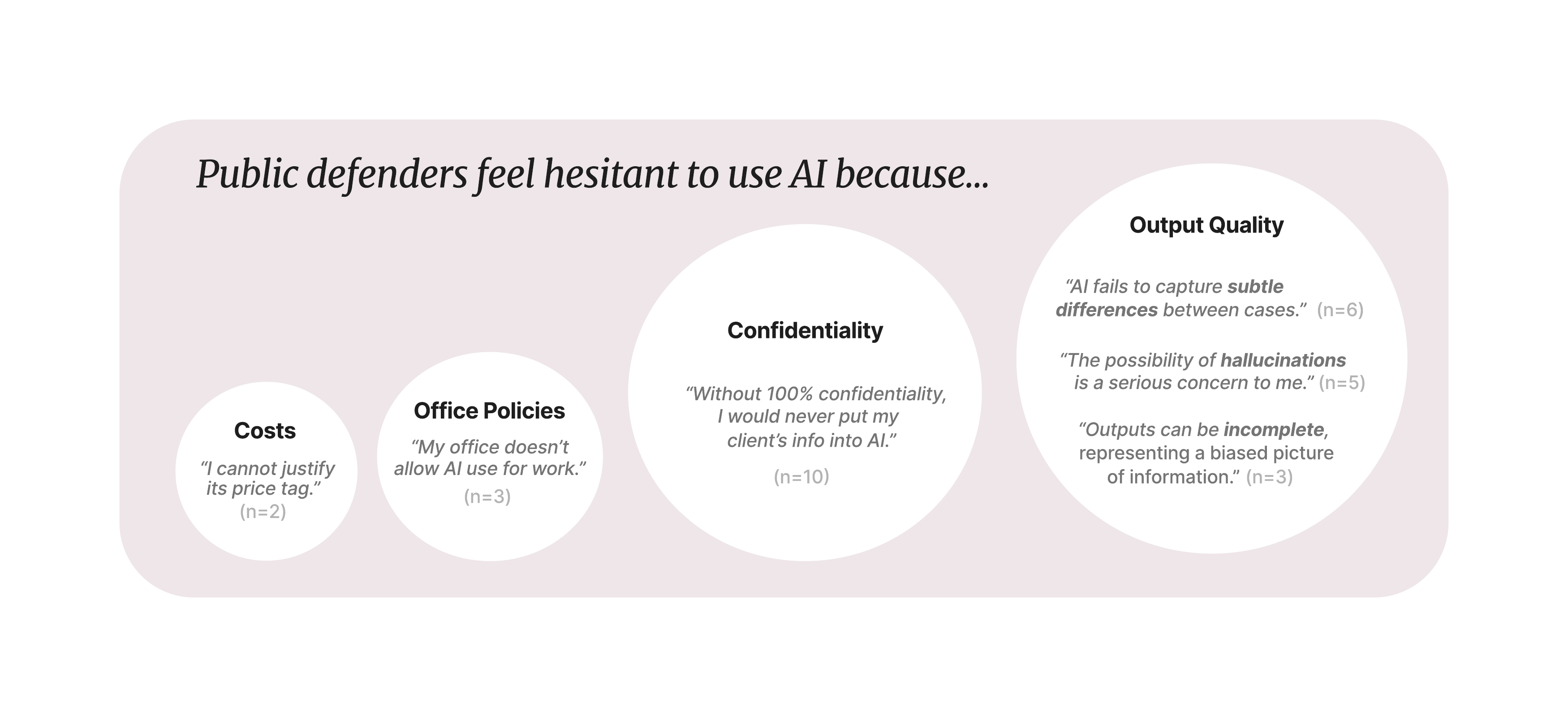}
    \caption{Main barriers to AI adoption in public defense work, grouped into four categories (costs, office policies, confidentiality, and output quality). The larger the circle, the more often it has been mentioned.}
    \label{fig:barriers}
\end{figure*}
 
\noindent\textbf{Costs.} Public defender offices and individual attorneys often lack the resources to experiment with or purchase commercial systems. P5 explained that all expenses must pass through county approval, a process that is both slow and uncertain: ``\textit{It’s expensive. I work for a public agency so every kind of new technology needs to be properly tested out and vetted, and the funding has to be approved. So it’s a little bit tricky that way.}'' The financial challenge is more acute for private attorneys who take on public defense cases without institutional backing. As P2 shared, ``\textit{I don’t have a Westlaw subscription just because I can’t afford it, and I’m not part of an office. The only legal database I have access to is the free one that comes with the Washington State Bar, and that’s Fastcase.}'' For these attorneys, even basic legal research requires navigating resource limitations, making AI services inaccessible. P14 explained that public defenders ``\textit{scrounge for investigation on [their] own budget}''. 

\vspace{5pt}
\noindent\textbf{Office norms and policies.} For public defenders employed by government offices, AI use is governed by institutional policies. Some offices approve enterprise-level systems only (``\textit{our offices said, stay away from it outside of Lexis or Westlaw}.'' (P1)), whereas in others no formal rules exist, and attorneys are guided only by broad cautionary advice. Private attorneys who take on public defense cases part-time are not bound by office policies and have greater freedom. 
 
Beyond workplace rules, participants noted the court warnings. P2 mentioned, ``\textit{I think some judges have issued rulings or standing orders that require attorneys to disclose if they've used any Generative AI.},'' but they found such orders ``\textit{vague and unenforceable because no one can really agree on a definition of what Generative AI means [such as whether it would include] a simple auto-correct or spelling mistakes.}'' Reflecting the growing attention and concerns around AI, more continuing legal education (CLE) programs (the ongoing professional education required for attorneys to maintain their licenses) offer courses on generative AI, but our participant (P7) reported that many trainings are not helpful as many of them focus on potential pitfalls, not applicable best practices.

\vspace{5pt}
\noindent\textbf{Confidentiality concerns.} Confidentiality is the foremost concern for all participants. Participants expressed that they would never submit any client-identifying information into AI systems, unless a system received formal approval from their office or underwent rigorous confidentiality clearance procedures. Participants expressed a strong preference for in-house AI tools that keep data within a ``\textit{closed universe}.'' As P7 explained, ``\textit{I would like a closed universe. I would like to be able to give as much information as possible to the tool to be effective, and confidentiality is what limits me from doing that}.'' Similarly, P13 was aware of several law firms purchasing ``\textit{closed versions of ChatGPT}.'' P5 underscored the challenge, noting that in most applications data ``\textit{necessarily has to go off site},'' processed on external servers, and that it ``\textit{would be nice if we could do it without taking any of that offsite}.'' Some participants also emphasized that protective orders explicitly restrict the use of AI. P3 and P4 explained that prosecutors provide unredacted sensitive discovery, such as bank records or warrants identifying confidential informants, only under agreements that this material will not be shared outside the office or submitted to AI systems. While these protective orders help public defenders access complete information, they also codify limits that make AI assistance unavailable, even for transcriptions.

\vspace{5pt}
\noindent\textbf{Unsatisfactory Output Quality.} More than half of participants (n=9) described current AI outputs as unhelpful or unreliable for their work. While tools could return surface-level matches, they frequently missed substantive legal issues. As P1 recalled:
``\textit{I was asking about the standard for ordering mistrial based on post-trial conduct, and it gave me a case in which there had been a mistrial in the guy’s first trial, but this decision was about a subsequent trial, so the issue of mistrial wasn’t part of what the case was about.}''
Similarly, participants noted that while tools produced technically accurate results for ``\textit{black-letter law},'' they lacked sensitivity to factual distinctions and the nuanced case-by-case analysis that characterizes skilled legal reasoning. As P5 explained:
``\textit{[i]t feels superficial to me. It feels as if the analysis is made on a very literal-minded level, and it doesn’t seem to really be able to read into those facts what the significance is of them, not what the definition is of the words, but what those facts mean in terms of the weight of the evidence.}''

Others emphasized that beyond obvious citation errors, more subtle hallucinations, such as misleading summaries or inaccurate interpretations, were more difficult to detect and could pose risks if unnoticed. P13 shared that ``\textit{[my] experience with the Westlaw and Lexisnexis [AI] is they'll produce text that sounds reasonable, but they often cite to cases that do not stand for the principles that they say the case says. [For example,] if you ask them a question about if a person has an insanity defense, will Missouri recognize temporary insanity, and it might give an answer that says the courts in Missouri are split on whether or not it will be recognized, but this court says it will if these conditions are met, and then they cite to a case. But when you read the case, the case does not say what the AI assistant said it says.}'' Despite this, some participants believed AI could accelerate certain tasks if used carefully. P2 described their experience: ``\textit{I asked [ChatGPT] ‘can you point me to the RCW, the Washington State Code,’ and it got it wrong the first time, but with [me] working it and babysitting it, [I] found [my] answer relatively quickly.}''

\subsection{What AI Can and Cannot Do: Five Pillars in Public Defense Work (RQ2)}\label{sec:pillars}

At the outset, we expected legal research and writing to be the main site where AI could play a role. As conversations unfolded, it became clear that public defenders’ work involves many more components, each presenting distinct opportunities and risks for technological support. Iteratively, we identified five broad categories that capture their everyday practice: (1) Evidence investigation, (2) Legal research and writing, (3) Courtroom representation, (4) Client communication and support, and (5) Defense strategy. For each, we distilled specific tasks from participants’ accounts and considered where AI assistance may prove most valuable.

\begin{figure*}[t!]
    \centering
    \includegraphics[width=0.96\linewidth, trim=1.2in 0in 1.2in 0in]{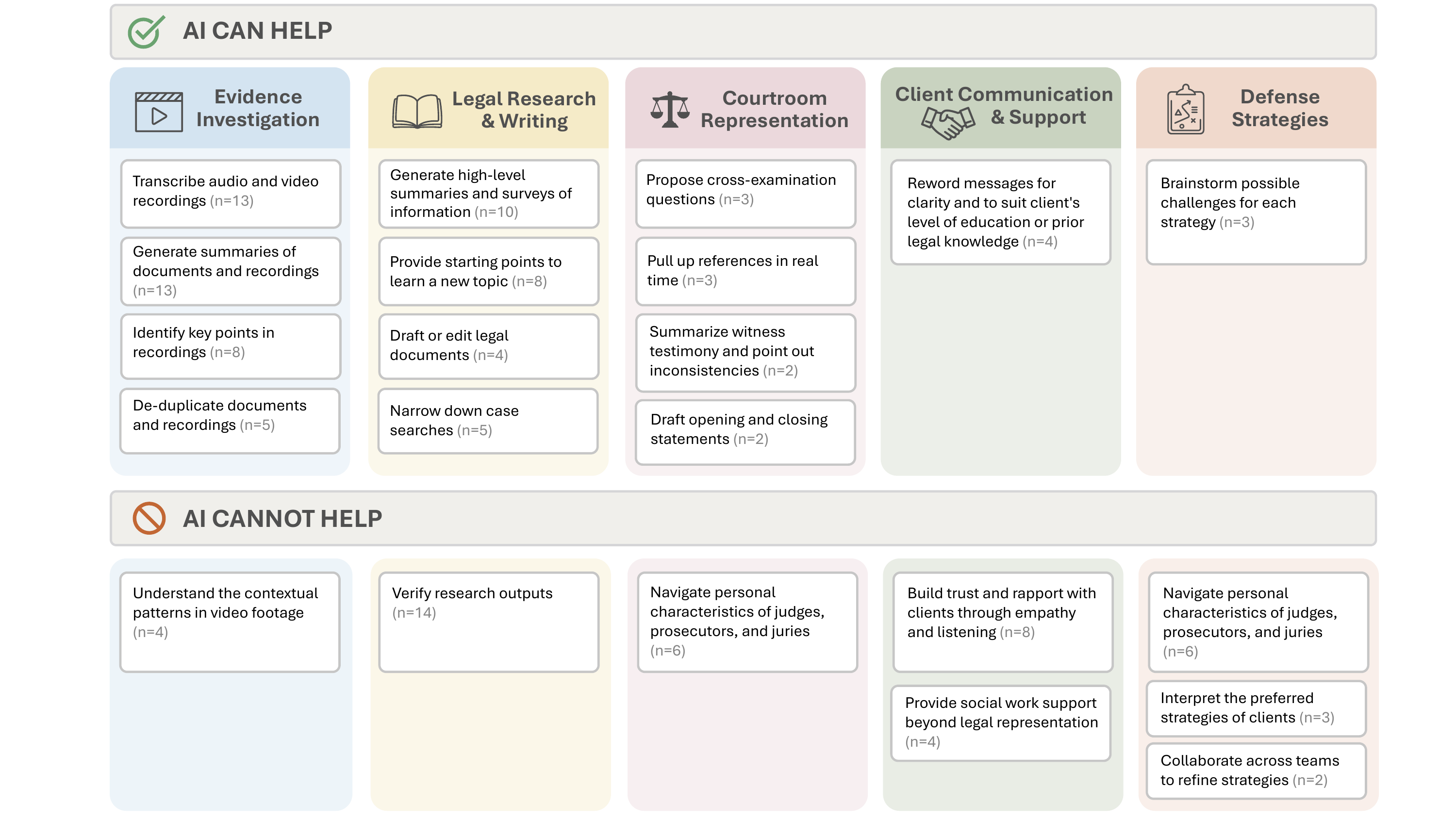}    
    \caption{Ways AI tools can and cannot help in public defense work. A majority of participants saw potential in evidence investigation and legal research and writing. Reading the personal characteristics of judges, prosecutors, and juries appears in two columns because participants raised it in two registers: as real-time judgment during courtroom representation, and as input to defense strategy.}
    \label{fig:Five_pillars}
\end{figure*}

\vspace{5pt}
\noindent\textbf{Evidence investigation.} Evidence investigation includes reading police reports, watching body camera footage, listening to audio recordings, as well as reviewing all other documents relevant to the case. Most participants noted that this step is \textbf{among the most time-consuming} in defense work, and could benefit heavily from AI assistance. They shared that it is often impossible to go through all the materials, due to the sheer volume and time constraints. The prevalence of body-worn cameras and phone calls usually leads to hours and hours of recordings that public defenders need to sift through. The recordings are informationally sparse, but it is difficult to know a priori which portion to watch or listen to. Police reports’ descriptions usually provide some pointers, but they might not be comprehensive either and may give a biased representation. 
Almost all participants (n=16) expressed that they would be \textbf{eager to use AI to help make sense of discovery data}. The most frequently mentioned tasks are transcribing videos or recordings and generating summaries. Defenders can search through transcriptions using keywords or parse them quickly: ``\textit{if only we could just send it all out for transcription, that would be a good start because I read a lot faster than I can hear.}''(P5) Similarly, summaries help triage whether a document is worth reading or a video is worth watching.

Participants also mentioned other methods of reducing the large volumes of data. Eight participants expressed interest in using AI for span extraction, which is identifying key segments in videos, such as asking the AI tools ``\textit{Can you give me the timestamp of when he reads the Miranda rights to the defendant}'' (P6), ``\textit{identify an area within a video [...] let's say warehouse door opens, and you don't want to go through six hours of video to find out when exactly that happens},'' (P2) and ``\textit{The crime occurs at [some location] at 4am, and there are some events that happen before that that might be pertinent to the event itself. The prosecutors turn over 12h of video, and probably 10 of those hours have nothing to do with the actual crime [...] AI could be used to identify the key parts of the video.}'' (P9) 
Five participants mentioned using AI to de-duplicate or piece together recordings and documents: ``\textit{I have a [...] case right now with over a hundred thousand records [...]. And there are so many duplicates. [...] And also be able to identify things that are like it's the same document.}''

Some offices already use AI tools for discovery review. Three participants (P5, P15, P16) shared that they use commercial AI tools for the tasks mentioned above in their office. However, while participants agreed AI could meaningfully assist with routine evidence review, they also emphasized that the \textbf{thought process during evidence investigation} is still guided by humans: ``\textit{it's not as if the program is doing the actual work, but it's framing things so that [we] know what point to fast forward to, and that just saved [us] two hours.}'' (P5) Additionally, they stressed that they would always verify the output of the AI tools. For example, they would search for key-words in the AI-generated transcripts, and would verify the timestamp of matched keywords in the original recording to ensure the transcript is accurate.

\vspace{5pt}
\noindent\textbf{Legal research and writing.} Legal research typically involves researching case law, statutes, or other legal and scientific materials, as well as drafting legal documents. Many participants (n=8) pointed to AI’s usefulness in generating summaries or \textbf{surveys of large volumes of legal information}. P1, an experienced research attorney, noted the difficulty of traditional Boolean searches, where ``\textit{95\% of the cases are going to be bad and 5\% good},'' and described winnowing as ``\textit{one of the most difficult tasks in criminal law research}.'' AI tools can speed up the process of identifying the most relevant information. Similarly, P10 envisioned AI’s potential in expansive multi-jurisdictional surveys: ``\textit{you can say, how does every State treat guilty except for insanity defenses [...] and it would go survey all the States [...] that could be really persuasive where you could say, our state is the outlier.}'' These reflections highlight how AI could accelerate tasks that defenders currently experience as laborious and time-consuming.
Participants also described turning to AI for quick entry points into new topics. Eight participants said they would use AI tools to quickly orient themselves: P2 reported that they often ``\textit{ask [ChatGPT] to generate quizzes to test [their] knowledge and retention},'' and P4 highlighted AI tools as ``\textit{a launching point for additional cases.}'' 

Participants also remarked that cases sometimes require research outside the law. As P11 explained: ``\textit{If [you] have a scientific area that’s relevant in [your] case, [you] need to become a mini-expert about that, and that’s very hard for a public defender to do with time constraints}.'' There are materials which are not indexed by WestLaw, such as finding relevant academic articles where AI could retrieve relevant documents. For example, P12 noted ``\textit{I recently used ChatGPT to find out if there were any law review articles on a certain topic that I was working on [...] and I tried ChatGPT, and it was really helpful because it searched Google}''. In such cases, participants believed AI’s ability to summarize and explain large bodies of information could be especially valuable. Notably, many participants said that legal research \& writing was a \textbf{relatively small portion} of their jobs. When drafting motions, they usually rely on templates or past motions, circulated in their office, instead of using AI. A few had experimented with AI to generate templates or edit drafts. For example, P11 found AI beneficial for editing writing: ``\textit{running your motion or your brief into ChatGPT, and say, edit this for clarity and concision you're gonna get, It's going to be an effective use, active tool.}'' (P11) But even in those cases, the bulk of the writing remained their own.

\vspace{5pt}
\noindent\textbf{Courtroom representation.} This category involves presenting arguments, examining witnesses, and persuading judges and juries. Most participants believed that AI could only play a limited role in this domain, though some saw potential for assistance in preparation. As P14 explained, ``\textit{Your cross-examination in court isn’t spontaneously conceived. If you’re a good lawyer, you are preparing for that in advance, and so that’s one other category that you could consider [AI adoption].}'' Others similarly imagined AI \textbf{supporting modular and auxiliary tasks} such as summarizing witness testimony, drafting opening or closing statements, or quickly pulling references (P5, P6, P11, P12). Looking ahead, a few participants speculated about more ambitious possibilities. P7 suggested that future tools might one day listen to oral arguments and provide real-time suggestions. But participants also highlighted why this is unlikely to be practical in the near term. As P5 emphasized, \textbf{courtroom tools would need near-perfect reliability}, since attorneys cannot stop mid-argument to verify an AI’s output. P8 argued that oral advocacy is central to the lawyer’s identity: ``\textit{That’s where the lawyer comes in.}'' Attorneys make emotional arguments and tell stories that resonate with human juries and judges, which, some participants think, AI cannot do. They also cautioned that the use of AI in court would depend on jurisdictional rules, since some prohibit even recordings or electronic devices in the courtroom.

\vspace{5pt}
\noindent\textbf{Client communication and support.}\label{sec:client communication} This category includes in-person and virtual interactions with clients to explain legal processes, share case updates, and discuss plea options versus trial risks. Ideally, defenders maintain regular communication with clients, though caseload pressures often interfere. Most participants mentioned AI tools could assist defenders in client communication, but would not be able to replace them. Some participants (n=4) already use or are open to using general-purpose AI tools to \textbf{reword messages for clarity} or to summarize documents for clients. P7 shared that they would use AI tools to help draft letters and emails to clients. They want to ensure ``\textit{[they] do not speak in legalese or convoluted ways, and ChatGPT helps keep [them] in check so that [their] parlance is a little bit more tailored to [their] audience.}'' P8 mentioned that his ``\textit{clients [...] have not gone that far in school. And so, one thing that you could use AI for is you could write a letter and then ask it, please, make edits to the letter to bring it to a 6th grade educational level.}'' Others imagined AI helping to create visual aids such as flowcharts or decision trees of the legal process that clients could keep as references after the meetings. P8 noted he often uses AI for ``\textit{massaging varying kinds of legalese writing into more common parlance description of what's happening. So you've got a very dense paragraph about the grand jury process and cross grand jury notice. And basically how this all works. And then using ChatGPT to boil that down to a more easily communicable idea of: you're allowed to testify at this proceeding, but we recommend that you don't.}'' 

Still, nearly all emphasized that \textbf{direct communication cannot be delegated entirely}. P11 noted, ``\textit{AI cannot substitute having a strong, healthy connection with [our] client and making [our] client feel like [we] care about them, and that [we’re] doing everything [we] can to help them}.'' P17 further emphasized that this human connection becomes especially critical when gathering sensitive personal information: \textit{most people want to have some kind of human-to-human connection, especially when you're asking deeply personal questions [...] They don't want to speak to AI or a computer in order to tell them their deepest, darkest secrets. You need to have some kind of personal level to extract that information.}'' Moreover, some participants, especially law professors leading criminal defense clinics, expressed concern that over-reliance on AI could prevent new defenders from developing crucial relationship-building skills. By outsourcing these interactions to AI, students would have fewer chances to develop such skills. Beyond the risk of over-reliance, participants also noted that clients' growing reliance on AI chatbots introduced a new layer of friction in attorney--client communication: ``\textit{clients come in with a lot of different things that they've gotten out of ChatGPT, and then we have to try to talk them out of that \ldots it does add another hurdle.}''  

Client support, however, extends well beyond legal communication. ``\textbf{Social work}'' or ``\textbf{holistic representation}'' emerged as a recurring theme in our interviews. Often, the root causes of a client's case extend far beyond what legal representation and courtroom advocacy can address, and more holistic representation models are gaining increasing recognition. As P14 explained, ``\textit{a good public defender does a lot of social work, [such as] getting their client into the drug program, getting their client mental health services, making sure their client gets enrolled in school, or does their community service to complete their diversion programs.}'' 

P15 stressed that the New Jersey Office of the Public Defender launched its holistic representation model in 2025~\citep{njopd2025holistic}, explicitly integrating legal representation with mental health services, housing support, and substance use treatment. P17 illustrated how social work affects sentencing outcomes: ``\textit{a client lived in a household where their father had routinely beat their mother, and they were subjected to violence [...] And what that does is allows the judge to take into consideration that they were a victim of domestic violence — [the judge] can sentence them a degree lower, so instead of 5 to 10 years, they would be sentenced in the 3 to 5 year range.}'' As addressing clients' needs holistically requires understanding the full complexity of their personal, medical, and social circumstances, participants did not identify meaningful use cases for AI in this domain. Instead, they called for greater recruitment of professional social workers.

\vspace{5pt}
\noindent\textbf{Defense strategy.} This category involves deliberate legal and tactical choices to protect clients against criminal charges. Public defenders develop case narratives that aim to exonerate, reduce culpability, or humanize the client, while also deciding whether to file suppression motions, challenge unreliable evidence, or negotiate pleas. A few participants imagined constructive roles of AI in strategizing. P6 suggested that AI might assist with ``\textit{brainstorming and also seeing how certain strategies play out… laying out the pros and cons of different options}.'' P7 envisioned a tool that could generate strategy suggestions and help identify weaknesses in the prosecution’s case to strengthen negotiations. 

However, because strategy blends procedural maneuvering, narrative building, and negotiation, participants also expressed skepticism about AI’s role. For many, strategic judgment was \textbf{inseparable from attorneys' lived experience}: ``\textit{I think I have more experience than ChatGPT does in representing clients}'', stated P14. It also involves the \textbf{preferences of clients}: ``\textit{a big part of a decision about strategy has to come from client, you don’t want to get the answers from AI}'' (P14). Both P10 and P14 stated that AI might have seemed useful early in their careers, before accumulated wisdom compressed the need for open-ended brainstorming. For example, P14 explained that misdemeanor clients in Washington, D.C. have almost no incentive to enter plea bargains, since the immediate penalties are minor and a guilty plea creates a lasting record, unless multiple convictions accumulate. In addition, P11 flagged the confidentiality concern of sharing the strategy with AI: ``\textit{Defense strategy is one of the most privileged pieces of information, and I wouldn’t want to just be putting it all in ChatGPT.}'' (P11) 

Beyond confidentiality, participants stressed that effective strategy requires both local relational knowledge and collaborative deliberation. P17 noted that \textit{if you practice in Atlantic County, you have a certain understanding of the prosecutors and the judges \ldots [That's] the humanistic aspect that we're always going to have, and that AI is not going to be able to replicate.}'' P15 echoed this, emphasizing that strategy development is fundamentally a \textbf{collaborative human process}: ``\textit{it's a very collaborative process [...] I might walk to 5 different offices and have 5 different conversations with my colleagues, trying to hash out that strategy [...] we're working with each other, we're collaborating, we're talking it through, we're all thinking of things that someone else didn't think of.}''

\subsection{Necessary Conditions to Foster Responsible Use (RQ3)}

\vspace{5pt}
\noindent\textbf{Mandatory Human Verification.}
All participants emphasized the importance of human verification. When using AI tools for legal and scientific research, P2 said that they would use the tools as a way to ``\textit{kickstart [their] research}'' and find the ``\textit{vocabulary necessary to do further research}'' with humans still steering the direction of the search. They would also always check the citations and references in AI outputs. Similarly, when using AI tools to assist evidence investigation, participants said that they would use AI for a first pass but still watch the recordings themselves to verify the accuracy of the output and to catch more things that AI tools might have missed, especially as the case progresses (P5, P7, P8). Looking ahead, P11 imagined that there needs to be a regulation in place to ensure that anything generated or modified by AI goes through a check conducted by humans because they think that people who are working under time constraints would be tempted to take AI outputs at face value. 

\vspace{5pt}
\noindent\textbf{Avoiding Over-reliance.}
Participants in teaching positions expressed concerns that over-reliance on AI tools does not help develop essential public defense skills such as efficient information gathering, effective communication, responding to unexpected situations, and decision making under pressure. First, participants noted that learning legal knowledge and skills from books is very different from engaging in public defense work in real life. For example, P12 noted that ``\textit{judges and prosecutors[, being humans,] may not act like what the court opinions make it seem like they do. If AI tools are modeled after what court opinions indicate would happen in a courtroom, it’s [going to] give [students] a false sense of what [they are] actually going to experience.}'' They further explained that an essential component of public defense skills is the ability to think on their feet and respond to unexpected situations in the moment. Since AI tools neither have the capability nor are allowed in the courtroom, public defenders will not be able to perfect everything they write through AI in the courtroom, but have to rely on their own abilities. Participants stated that although AI tools can assist legal research, it is still important for the public defenders to perform their own research outside of the AI generated responses. AI responses may not be comprehensive and may be a biased representation.
As P11 noted, if law school students and early career lawyers become dependent on having AI outputs, ``\textit{are we going to create a whole generation of lawyers and students and just people who have lost the ability to think creatively on their own? AI makes mistakes, and it’s important to make sure that our critical thinking stays sharp.}''

\vspace{5pt}
\noindent\textbf{Valuing Human Relationships and Judgments.}
Five participants emphasized that certain aspects of public defense rely on human relationships and human judgment, and therefore cannot be delegated to AI. Unlike technical tasks, these activities draw on contextual awareness, accumulated experience, and the relational value of time and presence. Participants explained that communication with clients depends not only on words, but also on non-verbal cues such as tone, posture, or hesitation. In the words of P10, ``\textit{the clients often give either explicit or more subtle feedback, and the students learn from that. They learned that there was a really awkward pause right there, the client reacted kind of harshly to that, and so they get a sense of maybe don't say that thing again, or maybe word it in a more careful way.}'' Participants stressed that these skills can only be learned from engaging directly with the clients and reflecting on these interactions.

Participants also described courtroom advocacy to be deeply tied to experiential knowledge. Public defenders learn preferences, temperaments, and informal habits of specific judges and prosecutors. As P12 put it, ``\textit{it’d be really hard for AI to help [lawyers] figure out what the best approach is in a given case because everything is also so personality specific.}'' Unlike human attorneys, AI systems do not have this accumulated background knowledge unless the user explicitly inputs it, and even then participants doubted AI can apply it with the same nuance. Participants further stressed that even if AI could one day observe such cues and generate contextual responses, clients might still feel undervalued: ``\textit{Clients have a right to counsel, not to machines.}'' (P14) For many, the time a public defender spends with them is itself a sign of care and commitment. This commitment is most concretely embodied in the social work and holistic representation discussed in Section~\ref{sec:pillars}. This practice dating to the 1970s that has demonstrated benefits for defendants' well-being, recidivism, and incarceration rates~\cite{matei2021assessing, buchanan2019impact}, but remains inaccessible to many offices due to resource constraints (P16).

\section{Discussion}\label{sec:05-discussions}

Our findings show that public defenders view AI as beneficial for the most time-consuming tasks such as evidence review, but they are constrained by confidentiality rules, cost barriers, and institutional policies. They also emphasize that strategy-building, courtroom advocacy, and client relationships are inseparable from human judgment, highlighting both the promise and limits of AI in practice. Building on these results, we turn to broader issues for the future: how to segment public defense work in ways that align AI adoption with professional values, how to move beyond a simple human–machine error dichotomy, and how open-source models can offer more secure and equitable alternatives.

\subsection{Future of Public Defense Work: Beyond AI Assistance in Legal Research \& Writing}\label{subsec:discussions-future}

Current discussions around AI use in the legal profession focus predominantly on generative AI for legal research \& writing~\cite{ABAopinion, tracker} driven by overall early excitement and concerns around LLMs \cite{NEURIPS2020_1457c0d6}, such as work stating that GPT-4 allegedly acing the bar exam \cite{katz2024gpt}. 
Some lawyers' irresponsible use of general-purpose LLMs for motion writing drew frustrated judicial warnings and media headlines about attorneys being disbarred. These incidents clouded broader perceptions of generative AI in legal work (\textit{see} Section~\ref{subsec:lit-concerns}). However, legal research \& writing are only one part of public defenders' practice. Public defenders spend a significant portion of their time distilling large records into concise summaries, classifying materials by relevance, synthesizing insights for strategic decisions, and preparing for cross-examinations. Such time-consuming tasks can be assisted by AI. For example, in video understanding, Video LLMs with advanced functionalities in classification, captioning, summarization, and question answering can support public defenders by enabling faster extraction of insights from complex audiovisual records~\cite{zhang2025videollama}. Table~\ref{tab:ai_pd_applications} maps some of these tasks to related streams of AI and NLP research.

\begin{table*}[t!]
\centering
\footnotesize
\caption{AI and NLP Technologies for Public Defender Tasks}
\label{tab:ai_pd_applications}
\begin{tabular}{p{2.8cm}|p{7cm}|p{6cm}}
\hline
\textbf{Task Category} & \textbf{Public Defender Need} & \textbf{Related AI and NLP Research} \\
\hline
Making Sense of Large Volumes of Data & 
Extracting relevant information from massive datasets, described as finding ``needles in the haystack.'' Tasks include summarization, span extraction, classifying data worth examining, identifying relevant documents or passages, and ranking data points by relevance. & 
Document classification \cite{cohen2006reducing}, passage assessment \cite{zheng2021doespretraininghelpassessing}, learning to rank \cite{learning_to_rank, guo2019deeplookneuralranking}, video span extraction \cite{grauman2022ego4dworld3000hours}, agentic web search \cite{deng2023mind2webgeneralistagentweb, openai2025operator, acharya2025agentic} \\
\hline
Legal Writing and Translating Legalese & 
Legal search, Drafting legal documents, translating legalese into accessible language for clients and target audiences. & 
Legal passage retrieval \cite{mahari-etal-2024-lepard}, legal analysis generation \cite{hou2024clercdatasetlegalcase}, translating legalese \cite{translating_legalese}  \\
\hline
AI as a Thought Partner/Educational Tool & 
Using AI for brainstorming, rapid learning of new topics, feedback on drafts, and preparation for cross-examinations or court hearings. Functions as an intellectual sparring partner for legal strategy development. & AI as thought partner/sparring partner applications \cite{collins2024building, 
vaccaro2024combinations}\\
\hline
\end{tabular}
\end{table*}

In line with Shao et al.~\cite{shao2025future}, we contend that AI/NLP research should be rooted in workers’ needs. Rather than framing the use of generative AI as a result of the laziness of attorneys, we argue for a proactive approach that incorporates advances in AI and NLP across different segments of public defense work. This requires detailed segmentation of tasks, since the ways of adopting AI, the associated risks, and potential mitigations vary widely. Blanket admonitions such as ``attorneys must verify all AI outputs'' offer little guidance. While checking citations in a motion might take two hours, verifying whether AI correctly flagged events in 200 hours of video is an entirely different burden. 

In addition, this segmentation must involve normative judgment on preserving the epistemic agency of human actors~\cite{malone2025trust, cheong2025epistemic}. Some components of work are essential to public defenders' efficacy (``\textit{In the courtroom, that’s where attorneys shine}'' (P8)), and some are essential to clients (``\textit{Clients have a right to counsel, not to machines}'' (P14)). Identifying this critical sphere of work is essential both for preserving professional identity and for designing healthier modes of human–AI interaction~\cite{coeckelbergh2015tragedy, sarkar2023enough}. For future research, we propose a four-step guide to structuring AI empowerment. 

\begin{enumerate}[nosep]
\item \textbf{Pain Points}: Which segment of work can be automated or augmented by AI? 
\item \textbf{Human Agency}: In each segment, what constitutes high-value work for focused attention? 
\item \textbf{Risks}: What risks arise from both the nature of the work and the technologies?
\item \textbf{Mitigations}: How can those risks be mitigated technically or organizationally? 
\end{enumerate}

\subsection{Beyond the Human v. Machine Error Dichotomy}

AI errors are inevitable just as human errors are. The important question is not whether they occur but which errors can be tolerated under what circumstances and how they can be controlled. Even advanced video LLMs can miss fine-grained details in blurry footage~\cite{zhang2025videollama}, and if AI misses decisive evidence during an initial sweep, public defenders may lose opportunities that could be critical for a case. Biases in many models also risk producing distorted narratives in sensitive contexts~\cite{tong2024eyes}. Human verification in such cases would require watching hours of footage in detail, undermining the purpose of AI adoption. With digital evidence expanding exponentially (what was once a single dash cam video has become twelve streams of footage from multiple angles), public defenders face a mounting burden that taxpayer-funded staffing cannot meet. Therefore, this challenge should not be reduced to a binary choice between human and machine fallibility~\cite{feng2025levels, shao2025future}. A more productive approach is to design systems that combine human and AI capacities so that fatigue-driven mistakes are mitigated and public defenders can devote more energy to strategy and client communication. The combination of low tolerance for errors, high human fatigue, and costly verification should not be treated as a dead end but as a research opportunity and design space.

\subsection{Commercial Platform Reliance in Legal Technology}

In our interviews, we noticed the pronounced presence of third-party vendors in this space. More than half of the participants mentioned specific commercial services or described hearing about contracts that other defender offices had signed. These tools are perceived as more secure than open platforms like ChatGPT, but more costly, requiring institutional subscriptions. Indeed, ``legal tech AI'' has become a lucrative business, with \$2 billion invested in startups in this field~\cite{legaltechstartup}. Scholars note that fragmented legal workflows (e.g., video analysis, docket management), combined with demands for specialized security guardrails, have enabled startups to secure contracts with large law firms~\cite{CodeXLegaltech}. Consolidation between vendors and big law is further fueled by the scarcity of legal data for training models. Although initiatives such as Free Law Project~\cite{CourtListner}, Caselaw Access Project~\cite{caselaw}, Pile of Law~\cite{henderson2022pile}, and Cambridge Law Corpus~\cite{ostling2023cambridge} have sought to democratize access, the core repositories of U.S. law remain behind costly paywalls. Startups therefore have strong incentives to partner with established firms to obtain domain-specific data~\cite{CodeXLegaltech}. 

This vendor reliance may deepen the justice gap by layering expensive AI services on top of an already proprietary information market, leaving under-resourced actors like public defenders further behind~\cite{bhambhoria2024evaluating}. The promise of confidentiality is also fragile. Vendors may promise encryption, non-retention, and local storage options. However, if these services rely on proprietary back-end models hosted by corporations such as Microsoft or OpenAI, it is unclear whether these systems can be considered a closed universe insulated from outside access. The integration of LLMs inevitably invites the risks of unintentional leaks or adversarial attacks~\cite{evertz2024whispers}. Moreover, the inputs and outputs handled by third-party vendors could fall outside of privileged information, subject to mandatory disclosure~\cite{privilege, cheong2024not, samaltman_confidentiality}. More importantly, there are no public evaluation benchmarks that allow public defender offices to audit third-party models whether they live up to their promised performance and confidentiality. Unchecked vendor dominance risks turning AI innovation into another barrier to justice rather than a bridge toward it.

\subsection{Pathways for Open-Source AI in Public Defense}

Concerns around vendor dependence make a strong case for utilizing open-source models~\cite{perron2025moving, bhambhoria2024evaluating}. Attorneys can run these models on-device, eliminating the risk of sensitive evidence leaving the public defender’s office, and the inputs and outputs of the systems are likely to be protected as attorneys' work product. In addition, open models allow more room for technical adjustments, such as citation retrieval systems, that make outputs more trustworthy~\cite{bhambhoria2024evaluating}. Recent open-source and multimodal models, e.g., Gemma 4 \citep{gemma4} or Qwen3.5 \citep{qwenteam2026qwen35omnitechnicalreport}, achieve remarkable benchmark performance and can now run on common laptops. They could be the base for in-house public defender office AI solutions: making sense of large volumes of evidence, or building search engines to retrieve relevant briefs, case alerts, best practices, and directives \citep{stammbach2026legalretrievalpublicdefenders}. Although local deployment requires upfront investment in hardware such as GPUs, it eliminates recurring expenses like subscription fees or API usage costs. The tasks public defenders identified as suitable for AI assistance in our study do not require the most advanced LLMs. What they need are practical, accessible tools that support high-volume, rule-based tasks like document review, triage, and structured evidence summarization and extraction.

To make open-source LLMs practically adaptable in public defense, two elements are essential: \textbf{in-house technical expertise} within public defender offices and \textbf{domain-specific LLM research} that fine-tunes models for the segmented tasks we discuss in Section~\ref{subsec:discussions-future}. For the first, states should consolidate resources to provide organizational support for public defenders. 
AI adoption will add new demands around data practices and the accumulation of institutional knowledge at scale. For the second, Perron et al.~\cite{perron2025moving} show that local LLMs achieve near-human accuracy in classifying and extracting substance problems in child welfare investigations. 
However, much work remains to bring such advances into public defense. High-quality datasets are needed, spanning legal texts, client communications, and long-form video data such as body-worn camera footage. 
Robust evaluation benchmarks are also critical, not only to measure text-level accuracy~\cite{magesh2025hallucination} but also to assess confidentiality, bias, and utility across various forms of data and contexts.

\section{Conclusion}\label{07-conclusions}

Public defenders carry a constitutional mandate under conditions of scarcity. Through qualitative interviews with public defenders, we examine whether AI can relieve parts of the burden. We categorize public defense work into five pillars, and we find that AI assistance is perceived to be most beneficial for time-consuming evidence investigation tasks. We identify three principles for responsible AI use by public defenders: human verification of AI output, awareness of over-reliance risks, and recognition that core aspects of public defense work rely on human relationships and judgment that AI cannot easily replicate.
We envision a future in which NLP and AI are responsibly integrated into public institutions, strengthening equitable access to essential services and supporting the people who deliver them. While there are no simple solutions, several priorities stand out. First, move beyond the human–machine dichotomy and design interfaces that make humans and AI more effective together. Second, direct research toward tasks central to evidence investigation, such as span extraction and video summarization, which public defenders identified as most practical. Third, invest in public defender–specific datasets and benchmarks to foster research and enable auditing of both open- and closed-source systems, including commercial tools. Finally, prioritize the development of locally deployable models that lower costs and safeguard confidentiality.

\clearpage
\bibliographystyle{plainnat}
\bibliography{references.bib}

\newpage
\beginsupplement
\section{Interview participants}
\label{appendix:participants}



\begin{table}[htbp]
\begin{center}
\begin{tabular}{|c|l|c|} 
 \hline
 \textbf{Participant} & \textbf{Affiliation / Current Role} & \textbf{Years of Experience} \\ [0.5ex] 
 \hline\hline
 P1* & Federal public defender & 40+ \\
 \hline
 P2* & Federal court-appointed private attorney & 7 \\
 \hline
 P3* & Federal investigator & 30+ \\
 \hline
 P4* & Federal public defender & 10 \\
 \hline
 P5  & State public defender & 20+ \\ 
 \hline
 P6  & Federal public defender & 10 \\
 \hline
 P7  & Federal court-appointed private attorney & 10+ \\
 \hline
 P8  & State public defender & 5 \\
 \hline
 P9  & Law professor, former public defender & 30+ \\
 \hline
 P10 & Law professor, former public defender & 2 \\
 \hline
 P11 & Retired law professor, former public defender & 30+ \\
 \hline
 P12 & Law professor, former public defender & 3 \\
 \hline
 P13 & Law professor, former public defender & 6 \\
 \hline
 P14 & Law professor, former public defender & 8 \\
 \hline
 P15 & State public defender & 20+ \\
 \hline
 P16 & State public defender & 2 \\
 \hline
 P17 & State public defender & 13 \\
 \hline
\end{tabular}
\end{center}
\caption{Description of participants.}
\end{table}

*P1 and P2 were interviewed jointly, and P3 and P4 were interviewed jointly.

\section{Legal Scenario}\label{app:scenario}
\begin{tcolorbox}[title=Case Example, width=\textwidth]
\small
\begin{description}[leftmargin=0pt, labelsep=5pt, itemsep=2pt]

\item[\textbf{Initial Stop:}] 
Officer Rodriguez observed a 2012 Honda Civic with a broken right tail light traveling northbound on Main Street in Bellevue, Washington. When Officer Rodriguez activated emergency lights to initiate a traffic stop, the vehicle initially slowed but then accelerated.

\item[\textbf{Pursuit \& Search:}] 
A pursuit began lasting approximately 5 minutes before the vehicle pulled into a gas station parking lot. Officer Rodriguez approached the car and claimed to detect the odor of marijuana. Based on this and the brief chase, officers conducted a warrantless search of the vehicle.

\item[\textbf{Evidence Found:}] 
50 grams of methamphetamine divided into 12 small plastic bags; Digital scale with residue; \$780 in cash in various denominations; Cell phone with text messages allegedly discussing drug transactions.

\item[\textbf{Defendant's Claims:}] 
The drugs belonged to a friend who borrowed his car earlier that day; Body camera footage has some audio gaps during crucial moments of the search; Recently attempted to enter a substance abuse program but was placed on a waiting list due to insurance issues.

\end{description}
\end{tcolorbox}

\section{Iterative Evolution of Interview Protocol}\label{} 

Our interview approach evolved from broad exploratory questions to targeted theoretical probes as emerging themes guided protocol refinement. This evolution exemplifies grounded theory's commitment to letting theory emerge from data through constant comparative analysis. 

\label{appendix:interview_evolution}

\subsection{Initial Interview Question Set}
\label{appendix:initial_protocol}

Our interviews began with a semi-structured protocol organized around the following core areas:

\paragraph{Prior Experience with AI Tools (10-15 minutes)}
\begin{itemize}
    \item How often, if at all, do you use AI tools inside and outside of work? Could you share what tools you have used?
    \item What do you think of using AI for work based on your prior experience and what do you know about these AI tools? What do you find promising or frustrating about these tools? What are you most concerned about regarding using AI tools?
    \item Which part of your work do you value the most? What are the parts that are time consuming but not as rewarding?
\end{itemize}

\paragraph{Case Deliberation (10 minutes)}
\begin{itemize}
    \item Can you walk us through how you would approach this case?
    \item Which step is usually the most time consuming or most difficult?
\end{itemize}

\paragraph{AI-assisted Information Retrieval (15 minutes)}
\begin{itemize}
    \item What service would you use first, and what search would you run?
    \item Follow-up questions based on participant responses and tool selections.
\end{itemize}

\paragraph{Thoughts on General AI Tools (10 minutes)}
\begin{itemize}
    \item Which parts of your work do you think is most likely to be automated, and would be most valuable to have AI assistance?
    \item What is missing in current AI tools?
    \item Which of these tools would be the most helpful for your work and how would you use them?
\end{itemize}

\paragraph{Closing Questions (5 minutes)}
\begin{itemize}
    \item Open-ended opportunity for participants to ask questions about AI, the research project, or comment on topics not covered in sufficient detail.
\end{itemize}

\subsection{Evolved Comprehensive Question Set}
\label{appendix:comprehensive_questions}

Through iterative refinement guided by emerging themes, our protocol expanded to include the following questions. However, these questions served as flexible guideposts rather than a rigid script—actual interviews varied considerably based on individual participants' experiences, interests, and responses, with new questions emerging organically during conversations and others being omitted when not relevant to particular participants' contexts.

\paragraph{Workflow Mapping \& Task Prioritization}
\begin{itemize}
    \item Looking at these five categories (Defense Strategy, Client Communication, Legal Research, Evidence Investigation, Courtroom Representation), do these five categories representative of your typical tasks? Do you want to edit anything?
    \item Where do you feel most overwhelmed in your caseload?
    \item What aspects of your work do you think we don't fully understand?
\end{itemize}

\paragraph{Case Strategy \& Decision-Making Process}
\begin{itemize}
    \item (Presenting the prepared case) When you get a new case like this case, what's your very first step? 
    \item (Presenting the five categories of work) How would you allocate your time across five categories of your task?
    \item Can you rank them from most to least time-consuming? What is usually the most challenging or difficult?
    \item How do you prioritize which issues to investigate first?
    \item What factors most influence your strategic decisions? How do you decide whether to pursue suppression vs. plea bargaining vs. trial?
\end{itemize}

\paragraph{AI Integration Boundaries \& Professional Identity}
\begin{itemize}
    \item Even if AI could help with various tasks, what work would you always want to do yourself? What aspects of your work feel too important to your role as an advocate to hand over to technology?
    \item What institutional barriers might prevent AI adoption in public defender offices?
    \item Do you think attitudes toward AI vary depending on roles (trial v. research attorneys), seniority (entry-level v. management), and employment types (private v. public)? What other factors could influence attitudes toward AI?
    \item What features would an ideal AI legal assistant have? What would most improve your ability to serve clients effectively?
\end{itemize}

\paragraph{Technology Assessment \& Tool Evaluation}
\textit{(Optional: when participants mentioned a specific tool)}
\begin{itemize}
    \item Is the output generated by ChatGPT and Westlaw AI helpful? What's missing or concerning?
    \item What improvements would make this more useful for your practice?
    \item What would make you more likely to adopt AI tools in your practice?
    \item What would you want AI developers to understand about public defense work?
\end{itemize}

\end{document}